# Turning the LHC Ring into a New Physics Search Machine

**Matti Kalliokoski,**[a] **Jerry W. Lämsä,**[b] **Mikael Mieskolainen,**[c] **Risto Orava**[c]

[a] *Beams Department, CERN, CH-1211 Geneva 23, Switzerland*
[b] *Iowa State University, IA 5011, U.S.A.*
[c] *University of Helsinki and Helsinki Institute of Physics, 00014 University of Helsinki, Finland*

*E-mail*: Risto.Orava@cern.ch

ABSTRACT: By combining the LHC Beam Loss Monitoring (BLM) system with the LHC experiments, a powerful search machine for new physics beyond the standard model can be realised. The pair of final state protons in the central production process, $pp \to p + X + p$, exit the LHC beam vacuum chamber at locations determined by their fractional momentum losses and will be detected by the BLM detectors. By mapping out the coincident pairs of the BLM identified proton candidates around the four LHC interaction regions, a scan for centrally produced particle states can be made independently of their decay modes.

# Contents

## 1. Introduction

### 1.1 Central exclusive production of heavy particle states

The Central Exclusive Production (CEP) of particle states, *X*, is described by:

$$pp \rightarrow p + X + p \qquad (1)$$

where the + signs indicate rapidity gaps. The CEP processes (1) can be produced by the initial states: (a) gluons ("double pomeron exchange"), (b) photons, or by (c) gluons and photons ("photoproduction" or "photon-pomeron" interaction). The respective cross sections for these processes (a-c) are calculated as the convolutions of the effective luminosities $L(gg^{\mathbb{P}\,\mathbb{P}})$, $L(\gamma\gamma)$, or $L(\gamma g^{\mathbb{P}})$, and the square of the matrix element of the corresponding subprocess [1,2].

The relatively small cross sections of the exclusive reactions (1a-c) are compensated by a number of advantageous properties compared to inclusive production.

- The masses and widths of the centrally produced *X*-particles are correlated with the fractional (longitudinal) momentum losses, $\xi_{1,2} = 1 - p_{f_{1,2}} / p_i$, of the final state protons ($f_{1,2}$) and the intial beam proton (*i*), as:

$$M_X^2 \approx \xi_1 \xi_2 s, \qquad (2)$$

where *s* is the centre-of-mass energy squared. A measurement of the invariant mass of the decay products would be required to match the *missing mass* condition available by the measurement of the pair of final state proton fractional momentum losses [3,4].

- In case of the Standard Model type Higgs-like decay branching ratios, CEP process (1a) is favoured since the leading order $b\bar{b}$ QCD background is suppressed by the *P*-even $J_z=0$ selection rule [3]. The exclusive CEP events are experimentally clean since the soft background is strongly suppressed, and a Higgs-like boson can be observed via the main decay mode $H \to b\bar{b}$. By studying the azimuthal angle distribution of the tagged leading protons, the quantum numbers of the central state (in particular, the *C*- and *P*-parities) can be analysed [3].

- At higher central masses, $M_X \geq 150\ GeV$, the photon-photon process dominates, and it is expected [5], that the excess of events at the γγ invariant mass of *750 GeV* recently seen by the ATLAS and CMS experiments [6,7] is generated by the photon-photon interactions.

**1.2 ATLAS and CMS di-photon state at 750 GeV as a bench-mark**

The recently observed excess in the di-photon invariant mass region around *750 GeV* by the ATLAS and CMS experiments at *13 TeV* represents some *N=20* events for the combined integrated luminosity of $L=5.8\ fb^{-1}$ [6-7]. The observed signal strength corresponds to a *13 TeV* inclusive production cross section of an *X*-particle decaying into a pair of photons of 6 - 10 fb, where a *40%* detection efficiency is given by the CMS collaboration [7].

For the year *2016,* an integrated luminosity of $L_{13TeV} = 26\text{-}31\ fb^{-1}$ is projected [8] yielding of the order of *200* inclusive events of the type registered by ATLAS and CMS experiments during Run 2 at *750 GeV*

**1.3 Present CEP tagging approaches at the LHC**

The CEP process (1) can be employed for particle searches provided that the CEP event candidates are triggered with sufficient efficiency and purity at the luminosities relevant for the search. For collecting the planned $\approx 30\ fb^{-1}$, an average pile-up rate of *50* events per bunch crossing is expected [8]. For the bench-mark particle production, a trigger efficiency of 40% would be required for reaching a total of 200 event candidates with a pair of final state photons.

With the event pile-up conditions foreseen at ATLAS and CMS experiments, the rapidity gap signature of process (1) has, in practice, survival probability of zero. It is, therefore, a major challenge to trigger on the CEP event candidates in the conventional manner. An estimated minimum time resolution of the order of 3 ps would be required[1] to separate the z-coordinates (along the beam direction) of the proton-proton interaction vertices [9]. No such performance is available for any of the leading proton detection systems deployed by the LHC experiments.

---

[1] To separate 50 interaction vertices within Δz = 5 cm luminous region a minimum distance sensitivity of 1 cm is assumed [8].

In the following, an alternative scenario that exploits the LHC machine infrastructure, is proposed for searches of CEP produced massive particle states.

## 2. Scanning for new massive particles

### 2.1 CEP protons exiting the LHC ring

The final state protons from CEP process (1), are injected into the LHC beam vacuum chamber and are subjected to the LHC beam optics conditions. In LHC collider's perspective, these protons represent "off-momentum" protons with momentum components that differ from those of the beam particles circulating along the optimal closed orbit. The CEP protons are eventually lost, either by the LHC cleaning-collimator system, or they exit the beam vacuum chamber at a distance, z, dictated by the fractional momentum loss $\xi$. Concerning the functioning of the LHC, these particle losses are of critical importance, and they are continuously monitored.

By tracing CEP protons of different $\xi$-values through the LHC accelerator lattice [10], a relation between the CEP proton exit points and the $\xi$-values of the final state protons is established. In the present study, only low-$p_t$ protons are considered as would be expected, for example, in photon-photon interactions (1b). Further study is being carried out for more general cases.

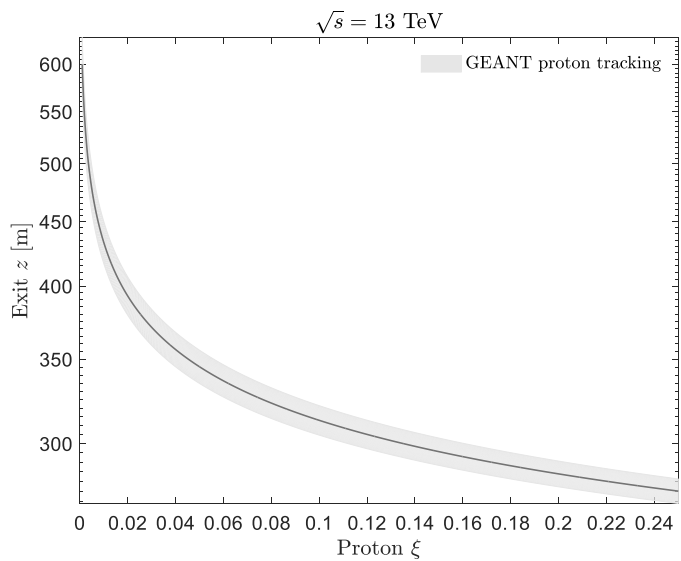

**Figure 1:** *The proton exit point, z, in CEP: pp → p + X + p, as a function of the fractional momentum loss, $\xi$ (solid line). The exit points of the leading protons out from the beam vacuum chamber are given in meters from IP5, the shaded band reflects smearing in proton transverse momentum[2].*

---

[2] The proton transverse momentum is smeared by *50 MeV*, which would be in accord with *dσ/dp$_t$∝exp(-20p$_t$)*.

In Figure 1, the proton exit points, shown as a function of their fractional momentum loss, $\xi$, are produced by the GEANT and proton tracing codes [10]. Through equation (2), the measured proton exit locations can then be used for an $M_X$ mass scan of the centrally produced systems (figure 2). The band widths reflect smearing in proton transverse momentum, $p_t$.

In principle, all CEP events producing massive new particles, independent of their decay modes, can be tagged by detecting the proton pairs exiting the LHC beam vacuum pipe. Assuming a branching ratio of $Br_\eta = 4\%$ for the hypothetical new particle state with mass of *750 GeV* [5], a total of *5000* events would be expected in 2016.

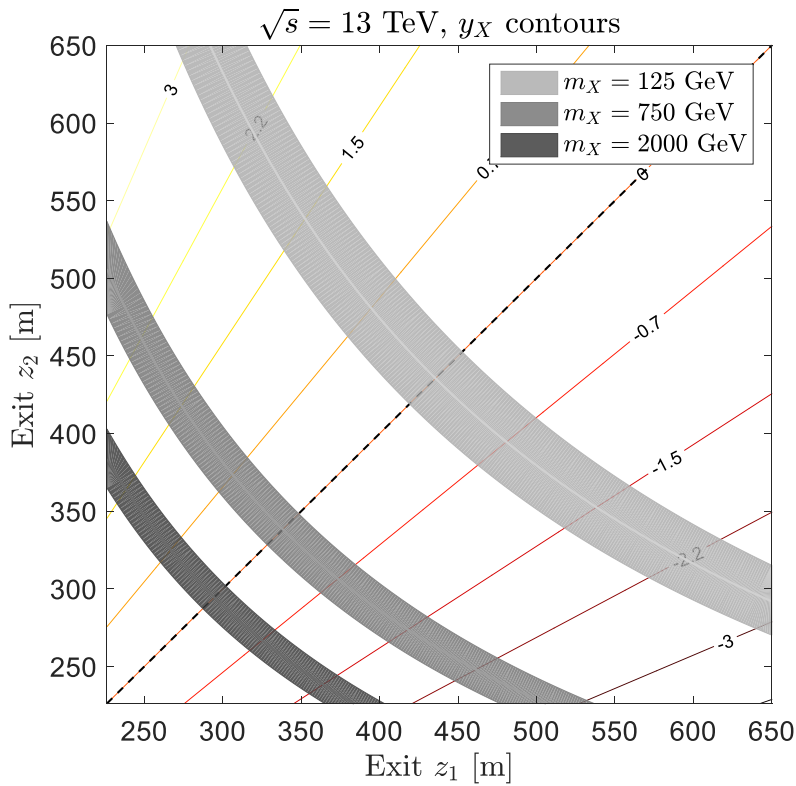

**Figure 2:** *The proton exit point combinations in CEP: pp → p + X + p, as a function of the central mass, $M_X$ (the grey bands). The exit points of the leading protons out from the beam vacuum chamber are given in meters from the Interaction Point 5 (IP5), the symmetric cases ($\xi_1 \approx \xi_2$) have $z_1 \approx z_2$ (dashed diagonal line). The rapidity span of the centrally produced decay products scales as $\Delta y \propto ln(M_X^2)$ (solid lines with the rapidity scale), rapidity of the centrally produced state is given as $y_X = 0.5ln(\xi_1/\xi_2)$.*

For the symmetric case ($\xi_1 \approx \xi_2$) in process pp → p + X(750 GeV) + p, a pair of *6125 GeV* final state protons exits the LHC beam vacuum chamber, and generates showers within the machine components. The showers generated by the exiting CEP protons can be described by the GEANT4 simulation package, where detailed CAD drawings are used for detailed descriptions of the LHC machine components. The azimuthal angular ($\phi$) distribution of the cores of the proton showers may be used to identify the interaction process (1a-c) in question [5].

– 4 –

**2.2 Tagging massive new particle states in CEP events**

By using the extensive BLM detector system around the LHC ring, massive particle states produced in the Central Exclusive Process can be detected independently of their decay modes. The following analysis approach is foreseen:

- Scan for the candidate CEP events by locating pairs of coincident proton exits on the opposite sides of the interaction point (IP) in question (figure 2).
- Correlate the tagged events with the LHC beam collisions (BCOs) within the time window for the chosen IP.
- Analyse the tagged LHC BCOs as candidates for CEP events with central masses, $M_X$, which would correspond to a registered pair of exit points (figure 2).

Depending on the experimental lay-out, all types of CEP processes (1a-c) could be detected.

**2.3 The LHC Beam Loss Monitoring System**

When the final state protons in the CEP process exit the beam vacuum chamber they will produce showers of secondary particles. These are detected by the Beam Loss Monitoring (BLM) system of the LHC. The LHC BLM system has almost 4000 detectors, mostly ionization chambers, spread around the ring. The main purpose of the system is to protect the machine components from critical beam losses by a beam abort when the measured dose in the chambers exceeds a threshold value [11].

In the BLM system, the signals from the detectors are digitized with a current to frequency converter; the pulses are counted over a period of 40 µs. The number of counts is passed every 40 µs to the surface electronics [12]. The surface electronics then combine the counts and creates integration windows by cascading multiple moving windows. In total, 12 time windows that span from 40 µs to 84 seconds are produced [13].

The recorded data are sampled by selecting the maximum value over one second and stored in a measurement database for a period of three months. These values are then resampled by storing only the ones which have changed from their previous readings [14].

**2.4 Further prospects**

The current LHC Beam Loss Monitoring system is specified for a continuous monitoring of potential risks to the machine operation. An improved time trigger is proposed to correlate the measured BLM loss to the event seen in each IP. As a straightforward upgrade to the present system, optical fibre connections are proposed for the locations of primary interest. In addition to providing a trigger for the CEP scans - independently of the decay modes[3] - the fibre connections facilitate independent Beam Loss Monitors, thereby improving the Machine Protection System (MPS) of the LHC.

Considering the different optical set-ups of the LHC interaction points, tailored BLM tagging based physics scenarios can be devised for each experiment. These will depend on the luminosities and detector lay-outs available, and can cover a wide range of standard model and beyond-the-standard model physics.

---

[3] Further studies concerning the invisible decay modes are ongoing.

## 3. Summary

The extensive radiation detector system of the LHC collider can be used to turn the LHC into a new physics search machine. By tagging the Central Exclusive Production processes, heavy particle states can be scanned as a function of their masses by detecting the exit points of the final state proton pairs. With minimal additions to the existing Beam Loss Monitoring system, an on-line trigger for new particle states could be provided.

## Acknowledgments

Albert De Roeck, Valery Khoze, James Pinfold, and Mikhail Ryskin are gratefully acknowledged for helpful discussions.

## References

[1] For pomeron-pomeron & photon-photon processes see, for example: V.A. Khoze, A.D. Martin and M.G. Ryskin, Eur.Phys. J. C23(2002)311; L.A. Harland-Lang, V.A. Khoze, M.G. Ryskin and W.J. Stirling, Int. J.Mod.Phys. A29(2014); Jerry W. Lämsä and R. Orava, 2009 *JINST* **4** P11019; M.G. Albrow et al., *FP420: A proposal to investigate the feasibility of installing proton tagging detectors in the 420-m region at the LHC,* CERN-LHCC-2005-025 (2005); *Forward physics and luminosity determination at LHC*, Proceedings, Workshop, Helsinki, Finland, October 31-November 4, 2000, ed. by K.,Huitu, R. Orava, S. Tapprogge, V. Khoze, 2001 - 226 pages, Prepared for 1st Workshop on Forward Physics and Luminosity Conference: C00-10-31.1 (New Jersey, USA: World Scientific (2001) 226 p.; The Helsinki Group: R. Orava with M. Battaglia et al., *Proposal to exted ATLAS for luminosity measurement and forward physics*, A Technical Report for the ATLAS Collaboration (2000) 93.7.

[2] For photon-pomeron processes see, for example: L.A.Harland-Lang, V.A. Khoze and M.G. Ryskin, Eur.Phys.J.C76, no. 1, 9 (2016)10.1140.

[3] V.A. Khoze, A.D. Martin and M.G. Ryskin, Eur.Phys.J.C19(2001)477; A.B. Kaidalov, V.A. Khoze, A.D. Martin and M.G. Ryskin, Eur.Phys.J.C31(2003)387A. DeRoeck, V.A. Khoze, A.D. Martin, R. Orava and M.G. Ryskin, Eur. Phys. J. C25 (2002) 391.

[4] J.W.Lämsä and R.Orava in *Diffraction at the LHC*, Rio de Janeiro, Feb 2002; R.Orava, *Forward physics measurements at the LHC*, Proc. of Science, DIFF2006:020 (2006); M.G.Albrow, T.D. Coughlin and J.R. Forshaw, *Central Exclusive Particle Production at High Energy Hadron Colliders*, Prog.Part.Nucl.Phys. 65(2010)149.

[5] L.A. Harland-Lang, V.A. Khoze, M.G. Ryskin, *The production of a diphoton resonance via photon-photon-fusion*, arXiv:1601.07187v1 [hep-ph] 26 January 2016.

[6] ATLAS Collaboration,"Search for resonances decaying to photon pairs in 3.2 fb$^{-1}$ of pp collisions at √s = 13 TeV with the ATLAS detector", Tech. Rep. ATLAS-CONF-2015-081, CERN, Geneva, 2015. http://cds.cern.ch/record/2114853.

[7] CMS Collaboration, *Search for new physics in high mass di-photon events in proton-proton collisions at 13 TeV*, Tech. Rep. CMS-PAS-EXO-15-004, CERN, Geneva, 2015. http://cds.cern.ch/record/2114808.

[8] J. Wenninger and J. Boyd in the LHC Performance Workshop, Chamonix, 25-28 January 2016, https://indico.cern.ch/event/448109/.

[9] M.G.Albrow et al., 2009 *JINST* **4** P10001.

[10] The off-momentum protons are simulated by using the following codes: MAD-X, GEANT3 and GEANT4 and Sixtrack by Jerry W. Lämsä; S. Redaelli, LHC MPS Review GEANT : R. Brun et al., GEANT3 Reference Manuel, DD/EE/84-1, CERN 1987; PHOJET: R. Engel and J. Ranft, and S. Roesler: *Hard diffraction in hadron-hadron interactions and in photoproduction*, Phys. Rev. D52 (1995) 1459: PYTHIA: T. Sjöstrand, P. Eden, C. Friberg, L. Lönnblad, G. Miu, S. Mrenna and E. Norrbin, Comput. Phys. Commun. 135, 238 (2001) 238.

[11] E.B. Holzer et al. *Beam Loss Monitoring for LHC Machine Protection,* Physics Procedia, Vol. 37, 2012, pp. 2055-2062.

[12] B. Dehning et al., *The LHC Beam Loss Measurement System,* in Proceedings of 22th PAC Conference, Albuquerque, New Mexico, USA, 25-29 Jun 2007.

[13] C. Zamantzas et al., *The LHC Beam Loss Monitoring System's Surface Building Installation*, in Proceedings of 12[th] Workshop on Electronics for LHC and future Experiments, Valencia, Spain, 25-29 Sep 2006.

[14] C. Roderick and R. Billen,*The LSA Database to Drive the Accelerator Settings,* 12[th] ICALEPS Int. Conf. on Accelerator & Large Expt. Physics Control Systems, Kobe, Japan, 12-16 Oct 2009.